# On a common carrier hypothesis for the 6613.6 and 6196.0 Å diffuse interstellar bands

R.J. Glinski  -  M.W. Eller

**Abstract**   We explore via spectroscopic modeling whether the highly correlated diffuse interstellar bands at 6613.6 and 6196.0 Å might originate from a single molecule.  Efforts were made to simulate the band contours of the DIBs along the three lines-of-sight, which have been observed by others at high resolution: HD179406, HD174165, and Her 36.  Reasonable simultaneous fits were obtained using a prolate symmetric top molecule that exhibits transitions of two different band types, type-a parallel and type-b perpendicular bands.  Two different excited states of a long- or heavy-chain, forked molecule are proposed.  A minimum number of adjustable parameters were used including ground and excited state $A$ and $B$ rotational constants, an excited state centrifugal distortion constant, and three different rotational excitation temperatures.  Points in favor and against the hypothesis are discussed.

## 1 Introduction

The long-standing problem of the identities of the carriers of the Diffuse Interstellar Bands (DIBs) has been well reviewed (Herbig 1975; Sarre 2006; Oka and McCall 2011).  Several hundred bands are confirmed as DIBs and the catalog continues to grow (Jenniskens and Désert 1994; Ehrendfreund et al. 1997; Hobbs et al. 2009).  It has long been known (Chlewicki et al. 1987;

R. J. Glinski and M. W. Eller
Department of Chemistry, Tennessee Tech University, Cookeville, TN, USA
email:  rglinski@tntech.edu

Krełowski and Walker 1987; Weselak et al. 2001) that some of the bands could be sorted into "families" based on the variations in their equivalent widths along lines of sight. The DIBs at 6613.6 and 6196.0 Å (here after called the λ6614 DIB and λ6196 DIB) have more rigorously been assigned to the same family (Friedman et al. 2011). In a robust statistical DIB correlation study (McCall et al. 2010), it was found that the correlation of these two bands with one another is "nearly perfect" along 144 diffuse cloud sightlines. The profiles of the two bands, however, are quite different and change differently depending on condition, which led researchers to question whether these bands could be from the same carrier (Oka et al. 2013). High-resolution spectra show that the λ6614 DIB manifests a broad multi-headed structure (Ehrenfreund and Foing 1996; Sarre et al. 1995; Galazutdinov et al. 2002) while the λ6196 DIB is narrow and relatively featureless (Galazutdinov et al. 2002; Kazmierczak et al. 2009; Kazmierczak et al. 2010).

It has been demonstrated that the carrier of many of the DIBs are most probably free flying molecules (ions or radicals) (Ehrenfreund and Foing 1996; Sarre et al. 1995; Kerr et al. 1988; Motylewski et al. 2000). Several attempts at spectroscopic analysis including rotational contour modeling have made plausible fits of the λ6614 DIB (Kerr et al. 1996; Schulz et al. 2000; Cami et al. 2004; Oka et al. 2013; Bernstein et al. 2015; Marshall et al. 2015) and the DIB near 5797Å (Edwards and Leach 1993; Rouan et al. 1997; Huang and Oka 2015). Dahlstrom et al. (2013) report that these two highly correlated DIBs show similar anomalous broadening in the spectrum of Herschel 36. In the work reported here, we have sought to model the highly correlated DIBs as being due to absorptions from the same ground state in a common molecule. We examine the constraints on the spectroscopic parameters that would define the hypothetical molecule and discuss the plausibility of whether two different excited states could belong to the same molecule. Molecular structures are suggested, which may be consistent with the parameters required of such a carrier; but a specific candidate is not proposed.

## 2 Description of the bands

### 2.1 The λ6614 DIB

This DIB is one of the most structured of all the bands (Galazutdinov et al. 2008a; 2008b). In the earliest high-resolution studies (Ehrenfreund and Foing 1996; Sarre et al. 1995), at least four band-heads are manifest; although workers have generally considered that only three heads are a part the band. Galazutdinov et al. (2002) have made high-resolution studies of this DIB along seven lines-of-sight, which show significant differences in the degree of excitation of the molecular carrier. Of the seven lines-of-sight studied, we have chosen to focus on the two lines that manifest the least and most band-head structure and broadening to the red. Recently, in an extensive DIB observing campaign, Dahlstrom et al. (2013) and Oka et al. (2013) have identified this band along the line-of-sight towards Herschel 36. In this case, the DIB exhibits "anomalous" behavior in that it broadens strongly to the red, particularly near the continuum; these "wings" were termed "extended tails toward red," ETRs. We present a summary of representative high-resolution spectra of the λ6614 DIB in Figure 1. The three lines-of-sight we have focused on in our modeling display three different degrees of excitation as demonstrated by the amount of broadening to the red. We refer to the lines-of-sight toward HD179406, HD174165, and Her 36 as the low-excitation, moderate-excitation, and high-excitation lines-of-sight, respectively. For the Her 36 spectrum, the observers have endeavored to isolate the contribution from the anomalously excited nebular material surrounding Her 36 from the contribution from foreground material; they have subtracted one-half of the DIB from a nearby star, 9 Sgr (Dahlstrom et al. 2013).

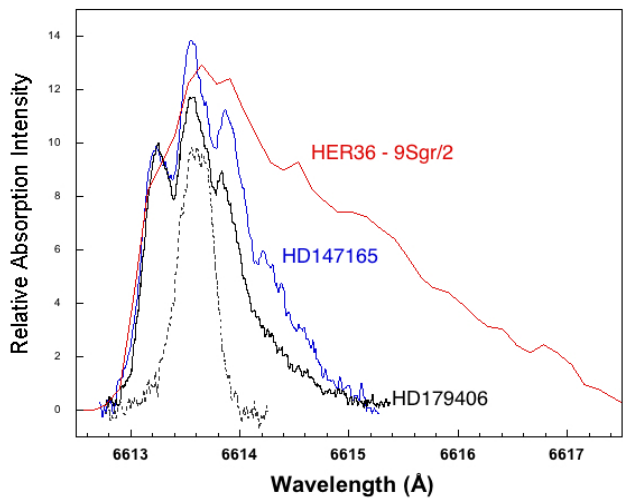

**Fig. 1** Three representative spectra of the λ6614 DIB: HD179406 and HD147165, Galazutdinov et al. (2002), R = 220,000; Her 36, Dahlstrom et al. (2013), R = 38,000. The λ6196 DIB is shown as the dotted line in the same frequency scale, to demonstrate the difference in the width of the two DIBs. Her 36-Sgr 9/2 spectrum represents the result of subtraction of one half of the intensity of the DIB obtained from Sgr 9 from the DIB in Her 36 (see text). Equivalent widths (McCall et al. 2010) are given for scale: HD179406, 104.9 mÅ; HD141765, 67.7 mÅ; and Her 36, 131.2 mÅ (along the entire line). Spectra are reproduced with permission.

A notable and important similarity in the two Galazutdinov et al. (2002) spectra is the congruence of the blue edge. We regard it as a strong spectroscopic constraint that the band does not broaden to the blue upon higher excitation. From our work and Galazutdinov et al. (2002), it is clear that this edge can be fit by the blue half of a Gaussian having FWHM of about 0.17 Å. We note that the blue edge also appears to be congruent in the separate spectra of Her 36 and 9 Sgr under the spectral resolution of that study (Dahlstrom et al. 2013; Oka et al. 2013), supporting the above observation.

Three clear band-heads are apparent in all seven of the Galazutdinov et al. (2002) spectra. In actuality, six of the seven lines-of-sight studied show four clear heads and five lines-of-sight show evidence of five heads. In the spectra shown in Figure 1, however, where five heads are identifiable in the spectrum of the moderate-excitation line, HD174165, only traces of added intensity is seen at the positions of the forth and fifth heads in the low-excitation line. Evidence of more than three heads in the Her 36 spectrum is weak; but it is unclear whether that is due to the lower resolution of the spectra or due to the velocity dispersions in the nebular environment. Most previous modeling has not endeavored to fit the fourth and fifth heads, while one (Bernstein et al. 2014) invokes a separate molecular species. Most recently, Marshall et al. (2015) make the case for vibrational hot bands and an arbitrarily imposed electronic component to fill the λ6614 DIB profile. Huang and Oka (2015) invoke spin-orbit states in order to describe the weaker heads on the DIB near 5797Å. For the purpose of this analysis, the shoulders, which are not perfectly resolved but seen reproducibly, are regarded as band-heads or subheads. We note that a very weak shoulder appears on the blue edge of the spectrum of Kerr et al. (1996) and on two or three of the lines-of-sight in Galazutdinov et al. (2002).

2.2 The λ6196 DIB

This DIB is one of the narrowest and is relatively featureless. Figure 2 shows the profile of the λ6196 DIB along the same three lines-of-sight as in Figure 1. The band broadens weakly, to both the red and the blue, particularly in the higher excitation spectra. (We note that the FWHM is not a good measure of the degree of shading of these DIBs as the broadening occurs near the baseline.) It can be seen in Figure 2 that a sharp head develops on the red side of the DIB and a lesser head to the blue; this trend is evident when examining all seven of the high-resolution spectra of Galazutdinov et al. (2002) and is seen even more clearly in Galazutdinov et al. (2008b) and Kazmierczak et al. (2009). Indeed, the appearance of the sharp red head on the λ6196 DIB correlates with the increased reddening and the appearance of the fourth and fifth heads on the λ6614 DIB. The features appear unresolved in the lower resolution spectrum of Dahlstrom et al. (2013); but, as with the λ6614 DIB, it is unknown whether this is due to instrumental or environmental reasons. Additionally, the DIB observed toward Her 36 also clearly shows a greater degree of excitation in the carrier, as the broadening is more pronounced especially near the continuum. The general properties of both bands are summarized in Table 1.

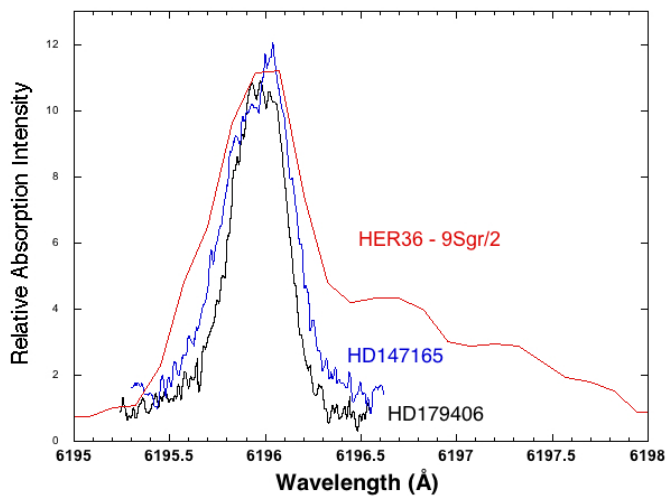

**Fig. 2** Representative spectra of the λ6196 DIB along the same three lines of sight as in Figure 1. Again, Her 36 is Her 36 – 9Sgr/2. Equivalent widths (McCall et al. 2010): HD179406, 21.5 mÅ; HD141765, 21.6 mÅ; Her36, 31.2 mÅ. Spectra are reproduced with permission.

**Table 1** Summary of the common carrier hypothesis

===================================================

| λ6196 DIB | State II |
|---|---|
| **Characteristics** | **Spectroscopic Constants** |
| Narrow. | $B_{II}' = 0.0071$ cm$^{-1}$ |
| One or two heads. | $A_{II}' = 0.29$ |
| Broadens weakly to blue and red, moderately to red in Her 36. | $D_{KII}' = 0.0$ |
| Parallel band form. | |

| λ6614 DIB | State I |
|---|---|
| **Characteristics** | **Spectroscopic Constants** |
| Broad. | $B_I' = 0.0060$ cm$^{-1}$ |
| Three to five heads. | $A_I' = 0.34$ |
| Broadens to red only, strongly in Her 36. | $D_{KI}' = 0.003$ |
| Perpendicular band form. | |

| | Ground State |
|---|---|
| | **Spectroscopic Constants** |
| | $B'' = 0.0070$ cm$^{-1}$ |
| | $A'' = 0.39$ |
| | $D_K'' = 0.0$ |

___

2.3 The nearly perfect correlation and common carrier hypothesis

These two DIBs have long been identified as belonging to the same family (Krełowski and Walker 1987) having been found to be closely correlated (Friedman et al. 2011). McCall et al. (2010) have found an R = 0.986 linear correlation coefficient between the equivalent widths of these two bands along 144 lines of sight. For comparison, those workers found an R = 0.985 correlation coefficient for two rovibronic lines originating from the same ground state level in the CH$^+$ ion along the same 144 lines. This is remarkable, as it requires a greater subjective judgment in striking a baseline under a 1.0-Å wide molecular band than under a 0.2-Å wide atomic line. Due to the different structure and behavior, researchers have reasonably dismissed the possibility that there was a common carrier; however, no detailed analysis was offered (Oka et al. 2013). Some evidence against the hypothesis has been reported by Krełowski et al. (2015); where the λ6196 DIB is observed in Orion OB1 association stars, while the λ6614 DIB is apparently absent.

The close correlation of the DIBs leads us to think that all of the heads and subheads belong inherently to each DIB as it would seem unlikely that any blended features would vary together. Next, we describe the spectroscopic modeling that seeks to test whether one molecular species can give rise to both bands. We will critically examine the nature of the excited states and consider plausible structures. Clearly, this is a complex spectroscopic problem and we regard this work as only a starting point.

## 3 Spectroscopic modelling approach

Given that the λ6614 DIB is greater than 2 Å (5 cm$^{-1}$) wide at the continuum and the λ6196 DIB is less than 1 Å (2.5 cm$^{-1}$) wide, it is difficult to see on first look how these features can be from the same molecule. Even more so, when considering that the bands would be generated by transitions from rotational ground states having one energy distribution. For this analysis we are taking it to be the case that both transitions share a common ground state. From a purely appearance point of view, the narrow band would seem to originate from a transition involving rotation of a rotor with a large moment of inertia. The broader band would seem to involve a transition involving a relatively small moment of inertia. A prolate symmetric top molecule that has very different $I_A$ and $I_B$ moments of inertia, and which also displays two different band types, might manifest both narrow and broad bands. Along these lines we seek to test the suggestion that two different electronic absorbance transitions can arise in one molecule.

3.1 A hypothetical prolate rotor having two different two band types

There are two features of a prolate rotor that might allow both of these two very different band profiles to appear simultaneously. A prolate molecule could: possess both a relatively large *A*-rotational constant but a very small *B*-rotational constant; and have electronic transitions that are termed type-a, parallel ($\parallel$) and type-b, perpendicular ($\perp$) band types (Hollas 1982).

Rotational energy levels in a prolate rotor are given by the expression:

$$F_v(J, K) = B_v J(J+1) + (A_v - B_v)K^2 - D_K K^4 \tag{1}$$

Centrifugal distortion constants beyond $D_K$ were found to only weakly effect the calculated profiles and were ignored. Using $A \gg B$ leads to a narrowly spaced stack of J states on top of a widely spaced set of K levels.

For a transition between singlet states, the two band types form under the following selection rules:

$$\text{For} \perp \text{bands:} \quad \Delta K = \pm 1 \text{ and } \Delta J = 0, \pm 1 \qquad (2)$$

$$\text{For} \parallel \text{bands:} \quad \Delta K = 0 \text{ and } \Delta J = 0, \pm 1 \text{ (but } \Delta J = \pm 1 \text{ only for } K = 0) \qquad (3)$$

The principle result of the two $\Delta K$ selection rules, when $A \gg B$, will be that parallel bands will pack most of the rotational lines into a narrow range of wavelengths and show only several tightly grouped band heads. The perpendicular bands will be more spread out and form heads spaced out approximately by twice the value of the $A$-rotational constant. Energy levels and transitions such as these are shown graphically in the classic books by Herzberg (1966; 1971).

We have constructed spectroscopic models using simple forms of a purely prolate molecule (where for the rotational constants: $A > B = C$), for singlet, non-degenerate vibronic states. (In all cases the Ray's asymmetry parameters were $\kappa < -0.99$.) We have sought models that best filled the extent of both the narrow and broad bands while filling the individual heads of each band. We let the ground state be defined by the rotational constants $A''$ and $B''$, with a rotational excitation corresponding to a single thermal temperature, T, for both the J and K stacks. The parameters for the two excited states are $A_I'$, $B_I'$ and $A_{II}'$, $B_{II}'$ for the λ6614 DIB and λ6196 DIB, respectively. Detailed spectroscopic modeling was accomplished using the program ASYROT (Judge 1987; Zuali 1994). The optimum spectroscopic constants obtained from the modeling are listed in Table 1. Reasonable fits could be accomplished using spectroscopic constants that are within about 5 to 10 % of those listed in Table 1.

The modeling approach was iterative. As a starting point, clues were discerned from the broadening. The simultaneous broadening to the red and weakly to the blue of the λ6196 DIB, suggests that $A_{II}' < A''$ and $B_{II}' \approx B''$. Broadening to the red only for the λ6614 DIB, suggests that $A_I' <$ or $= A''$ and $B_I' < B''$. Following the hypothesis that the λ6614 DIB is a perpendicular band, we began with an $A''$ rotational constant of about 0.40 cm$^{-1}$, which dictated a rotational excitation temperature near the cosmic microwave background (CMB), 2.7 K, consistent with that of interstellar CN (Crane et al. 1986) and the results of Huang and Oka (2015) in their modeling of the DIB near 5797Å. A qualitative fit of the multi-headed λ6614 DIB using various $B''$ constants was obtained, so attention was turned to the λ6196 DIB; the narrowness of this DIB is the stronger constraint on the values of the *B*-constants. The Gaussian line-width parameter was chosen to fit the blue edge of the λ6614 DIB, 0.17 Å. A slightly broader line-width parameter was used to fit the Her 36 DIB, 0.21 Å. As the modeling progressed we found it useful to add one more parameter, the $D_{KI}'$ centrifugal distortion constant, in order to fit the blue edge of the λ6614 DIB. Alternatively, we found this also could be accomplished if $A_I'$ is much less than $A''$. Reasonable fits were accomplished with $D_{KI}'$ ranging from 0.001 to 0.005 cm$^{-1}$, with corresponding values of $A'' - A_I'$. The calculated band contours were arbitrarily shifted to align with each DIB (but not shifted with respect one another).

    To summarize the modeling approach, these are the assumptions made in this modeling:

1. The close correlation of the DIBs is not coincidental and the bands represent absorptions from the same ground state of one molecular species, likely into two different electronic excited states.

2. The band structure is as described in the text and summarized Table 1, where all of the subheads are regarded to belong to each DIB.

3. A pure prolate structure is assumed that has large *A*-rotational constants and very small *B*-rotational constants.

4. The molecule displays both parallel and perpendicular rovibronic band types; only singlet states could be handled by the models.

5. The $T_J$ and $T_K$ are assumed to be the same thermal (Boltzmann) temperature, the lowest one set at the value of the CMB.

The simplest phenomenological model was sought, using the fewest possible number of adjustable parameters. No other specifics of the molecule are assumed, allowing us to discuss whether the required spectroscopic constants can be plausibly proposed. The manifestation of these two transition types in one molecule is discussed below.

**4 Results of the modeling**

4.1 Simultaneous fits of the two bands

Spectra were calculated using the parameters in Table 1 in effort to model the contours of the two bands simultaneously. Perpendicular and parallel band profiles were calculated at three different temperatures to model the λλ6614 and 6196 DIBs, respectively. The results are shown in Figures 3 and 4, together with published spectra along the three lines-of-sight that display three degrees of rotational excitation.

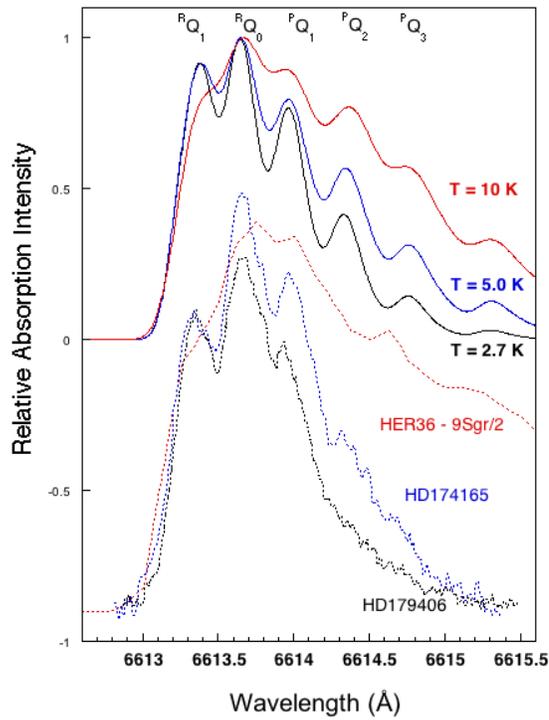

**Fig. 3** Modelled spectra at three temperatures are shown as solid lines; DIB spectra are shown as dotted lines and are the same as those in Figure 1.

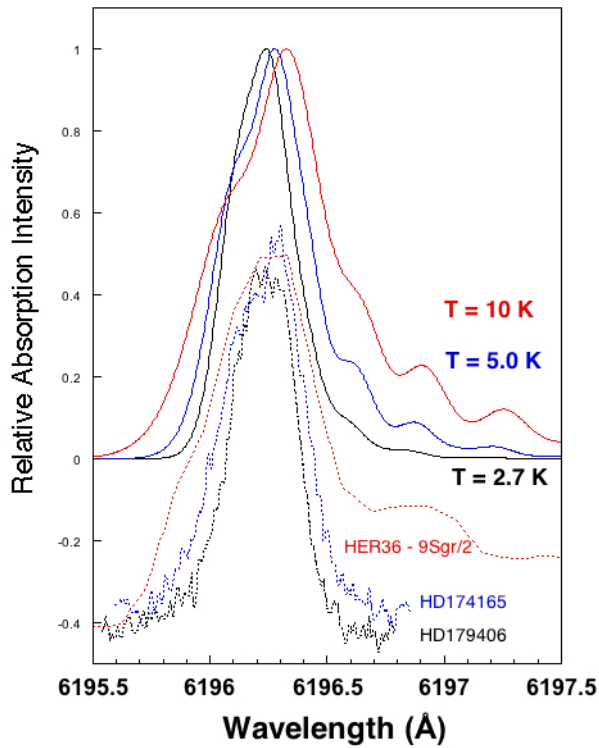

**Fig. 4** Modelled spectra at three temperatures. DIB spectra are the same as those in Figure 2.

The fits of both DIBs are quantitative with respect to the widths of the DIBs, the band-head spacing, and the broadening trends; while the general appearance of the band-heads is qualitatively described with respect to the relative intensities. The fit of the ETR of the λ6614 DIB in the Her 36 line is probably as good as use of a single thermal temperature would allow. Additionally, the models are successful in describing the broadening to both the blue and the red in the λ6196 DIB, as well as the "wing" formation toward the red; although this is not perfectly quantitative. Addition of a higher temperature component to the 10 K model, could improve the fits of the wings of both bands. We note that the prediction that the λ6196 DIB will form a red head is qualitatively shown, but the lower resolution of the Her 36 spectrum prevents clear determination of the trend. The slight drift to the red of the peak maximum of the λ6196 DIB predicted by the modeling is not clearly seen in the spectra. As described above, the strong match of the blue wing in the λ6614 DIB models was forced by choice of line-spread function and use of a relatively large D$_K$ constant. Alternatively, choice of a large $A'' - A_1'$ would also force a turn around of the $^RQ_K$ heads at the blue edge.

As can be seen in Figure 3, a significant exception to the quality of the modeling is that appearance of the fourth and fifth heads in the low-excitation spectrum of the λ6614 DIB is over-predicted by the model. Some extra intensity is evident at the two positions, but there are no clear heads. Additionally, some strength differences are not accurately predicted for the three main heads by our simplest models. Several factors may contribute to this. We have considered thermal temperatures that are the same for both the K and J rotational energy stacks. Since it is likely that the excitation mechanism is not thermal but radiative (Oka et al. 2013), it will be likely that the K and J temperatures will different. Furthermore, the distributions being non-LTE will lead to population distributions that are not reflective of one temperature; instead, there will be a tail of highly excited species in the distribution; i. e., it would enhance the formation of broad "wings" at the continuum. We also note that each of the three main heads are not fully filled in the spectrum of the λ6614 DIB. This suggests that the J-temperature may be greater than the K-temperature, at least in the low excitation spectra. The combination of these factors and an increase in the J-temperature would broaden the heads individually; thus flattening the weaker heads. We note that the K-temperature might possibly be less that 2.7 K; this has been observed in microwave spectra of formaldehyde in regions where radiative pumping sets the rotational energy distributions (Palmer et al. 1969; Townes and Cheung 1969; Thaddeus 1972). We obtained better fits at temperatures less than 2 K, but we feel this is not defensible with the current state of knowledge.

We have not explored what the spectrum from an asymmetric rotor would look like nor have we considered hybrid bands (Herzberg 1966) since we did not wish to add to the number of adjustable parameters. Additionally, we have only considered singlet states and have ignored Coriolis interactions. We think that the above modifications would tend to "smear" the features of the perpendicular band. The spectra may be made more complex by the presence of hydrogen atoms, which may add nuclear spin statistics to the relative intensities of the band heads. The inclusion of a more sophisticated spectroscopic model, including non-LTE rotational energy distributions would be a complex undertaking, for which there is currently not enough data, and is left for future work. We have specifically not used a myriad number of adjustable parameters to explore a number of spectra; nor have we invoked spin-orbit states or blends of spectra from different carriers.

We have shown that it may be possibly that the two DIBs could be due to transitions from the same ground state, but the two different excited states would have to have relatively large geometry differences.

## 5 Discussion

The purpose of this study was to explore whether the transitions corresponding to the two highly correlated DIBs could begin in the same ground state of one molecule. Specifically, could the transitions yield a broad, structured band and a narrow, relatively featureless band. We have shown that this may be possible and discuss various questions concerning the result.

### 5.1 Is it possible for one molecular species (ion or radical) to manifest both a parallel and a perpendicular band in its electronic absorption spectrum?

For the hypothetical molecule considered here, the geometries of the two proposed states that fit the DIBs are quite different. Hence, the two transitions in the candidate molecule would have to be from two different electronic excited states. Therefore, we make a preliminary inquiry into what these states may be like. The rules for spectroscopic transitions are guided by both the symmetry of the electronic and vibration state involved. For the purpose of discussion, we will consider the electronic spectrum of a candidate prolate molecule with symmetry corresponding to the $C_{2v}$ point group. The vibrationless ground state has the irreducible representation, $A_1$. Let us assume that the lowest excited electronic state has orbital symmetry, $A_2$. In order for a transition to be electric dipole allowed, the direct product of the irreducible representation of the two states must correspond to a representation of one of the dipole axes. The vibrationless transition $A_2 \leftarrow A_1$ is forbidden as an electric dipole transition (the spectroscopic convention is used here where the excited state is given first). If, however, the electronic transition is accompanied by a vibrational excitation, the upper "vibronic" state takes the symmetry of the direct product of the symmetries of the electronic state and the vibrational mode. So, for transition corresponding to $A_2 \leftarrow A_1$ with an out-of-plane vibration having $B_1$ symmetry, the upper vibronic state will have the overall symmetry, $A_2 \otimes B_1 = B_2$. The direct product of the ground state and the state symmetries will then be, $A_1 \otimes B_2 = B_2$; which will correspond to an allowed transition polarized perpendicular to the rotational axis (Steinfeld 1985).

Therefore, the state called State I in Table 1 could have $A_2$ symmetry and yield a vibronic transition showing perpendicular polarization. State II, could be a higher energy state with symmetry $A_1$, having electric dipole allowed transition of parallel polarization. These rules for the nature of the vibronic transitions must apply to any molecule in the $C_{2v}$ point group. We have not considered the effects of molecules having $C_s$ symmetry, nor have we allowed for hybrid bands that have both type-a and type-b characteristics.

A concrete example of this is formaldehyde, $CH_2O$, a well-studied molecule that displays both parallel and perpendicular bands in its electronic absorption spectrum (Moule and Walsh 1975; Moule and Judge 2003). For $CH_2O$, however, both bands occur in the same $^1A_2 \leftarrow {}^1A_1$ transition, which is relatively weak. The vibrationless absorption is observed as a magnetic dipole transition and displays a parallel band that has relatively narrow profile at low resolution. Vibronic transitions that occur with an odd-number change in the $\nu_4$ vibrational quantum number show up as much broader perpendicular bands.

5.2 Are the spectroscopic constants required for these fits plausible
   for a molecule and what type of molecule?

It would take a special structure to be purely prolate (or nearly so) and have such a large difference in rotational constants: $A >> B = C$. We suggest there is general structure of such a molecule and have found two different molecular structures, which possess such properties; these are shown in Figure 5. The most general "skeleton" structure would be a forked chain. We found that the rotational constants required for our modeling, given in Table 1, would correspond to either of the two structures 5b and 5c, using the specific geometries listed in Table 2. None of these structures are being proposed as a carrier; they are only presented for discussion.

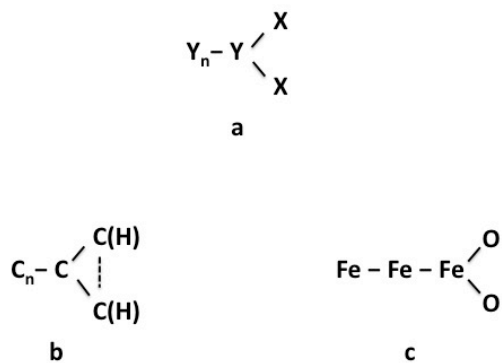

**Fig. 5** Prolate structures consistent with modeled species that lead to the calculated spectra. The trigonal structure may or may not be cyclic.

**Table 2** Geometries of the hypothetical skeleton structures in Figure 5 that lead to the rotational constants in Table 1.

|  | $R_{yy}$ | $R_{yx}$ | $\angle xyx$ |
|---|---|---|---|
| **Structure b (n = 10)** | | | |
| State II | 1.29 Å | 1.55 Å | 150° |
| State I | 1.40 | 1.45 | 145 |
| Gnd. State | 1.27 | 1.40 | 134 |
| **Structure c** | | | |
| State II | 2.65 | 1.70 | 105 |
| State I | 2.85 | 1.70 | 95 |
| Gnd. State | 2.67 | 1.57 | 96 |

Note: All bond lengths in the chains are considered equal for each structure.

The carbon containing structure, 5b, is much like the polymethene chain species discussed previously (Kroto et al. 1987; Jochnowitz and Maier 2008). The possible presence of the cyclopropenyl group was explicitly discussed by Thaddeus et al. (1993). The structure is not inconsistent with the kind of species discussed by Huang and Oka (2015), with the exception that structure 5b is a bit smaller than they prescribe. The longer chain, however, could bring the absorption spectrum into optical wavelengths (Maier and Fischer 1997; Maier 1998). The carbon chain may be terminated on the left by another atom, e.g. nitrogen or oxygen (Blanksby et al. 2001), in order to provide a large dipole moment, which has been shown to be necessary to establish the radiative equilibrium by Cami et al. (2004) and by Huang and Oka (2015). The relatively large change in the $A$-rotational constant is relatively easy to account for with these structures due to the fact that a small change in the XYX-bond angle will lead to a large change in $A$. The large change in the $B$-rotational constant might be considered more problematic, especially since the bond lengths in the polymethene chains are calculated to not change greatly on electronic excitation (Fischer and Maier 1997). We note that a relatively large change in $B$ was needed to fill the profile of the λ6614 DIB by Oka et al (2013). For reference, the CC triple bond distance in acetylene is 1.21 Å, the CC double bond distance in allene ($CH_2=C=CH_2$) is 1.31 Å, and the CC single bond in ethane is 1.54 Å (Herzberg 1966).

If it is unlikely that these large geometry changes can be accommodated by the carbon-chain structure in Figure 5; it is plausible that a smaller molecule, like the structure in Figure 5c, could show those geometry changes between the several excited states. With regard to the iron-containing structure, Figure 5c, we have no basis for thinking this species exists--it is merely considered as it satisfies the criteria of yielding the requisite rotational constants. Although, where as fragments of carbonaceous grains have generally been considered as DIB carrier candidates, fragments of other types of grains have not been considered. The bond distances $r_{FeFe}$ are estimated from the computational studies on iron clusters (Pauling 1976; Gutsev and Bauschslicher 2003) and the $r_{FeO}$ distance is consistent with that of FeO (Huber and Herzberg (1979). Nothing in this work constrains the identities of the X or Y atoms in Figure 5a, except that the chain is heavy and/or long and that the X atoms must have masses such as that of a carbon, nitrogen, or oxygen atom. Other structures could be proposed containing magnesium atoms (Ziurys et al. 1995) or aluminum atoms (Jochnowitz and Maier 2008). Finally, we note that a general $CH_2X$ structure has been suggested to account for regular, distinguishing, and testable (Tada et al. 2006) forms in the rotational and vibronic structure of a small subset of DIBs (Glinski and Nuth 1995; Schulz et al. 2000).

Previous workers have demonstrated that any transition describing the DIBs must have a relatively high oscillator strength (Huang and Oka 2015). We have made no attempt to take into account the absolute intensities of the bands; i. e., we have not considered the transition moments for the bands, calculated equivalent widths, or predicted column densities as this was beyond the scope of our work.

5.3 What can be said about a DIB identification strategy?

Recent robust DIB observations, including high-resolution spectroscopy studies, and the discovery of anomalously high excitation lines of sight are bringing about new optimism for DIB identification. Further spectroscopic contour modeling, such as this work and the work of Bernstein et al. (2015) and of Huang and Oka (2015), could yield new insight into this problem. The relative dearth of molecular and spectroscopic data, however, especially on molecules similar to Figure 5c, makes it difficult to rigorously test the common carrier hypothesis, or any proposed identification. There is continuing need for concerted laboratory spectroscopy programs (Schmidt and Sharp 2005) and high-level theory calculations on exotic molecules—ones not necessarily containing carbon. We conclude that the common carrier hypothesis is supported by several results of this work and that further explorations along these lines are encouraged. But there is considerable work left to be done before any of these results may be called a breakthrough in the identification of any of the diffuse interstellar bands.

We thank Erik Hoy for help with the early calculations and David Dixon for references to the iron cluster data. The DIB spectra from HD179406 and HD147165 are reproduced with permission from Jacob Krełowski; that from 36 Her is reproduced with permission from Julie Dahlstrom.